\documentclass[aps,prl,twocolumn,showpacs,superscriptaddress,groupedaddress,nofootinbib,nobibnotes]{revtex4-1}

\usepackage{breqn}
\usepackage{graphicx}
\usepackage{dcolumn}   
\usepackage{bm}        
\usepackage{amssymb}   

\usepackage[breaklinks=true,colorlinks=true,linkcolor=blue,urlcolor=blue,citecolor=blue]{hyperref}

\usepackage{float}

\begin{document}

\title{Electromagnetic turbulence suppression by energetic particle driven modes}

\author{A.~Di Siena} \affiliation{Max Planck Institute for Plasma Physics Boltzmannstr 2 85748 Garching Germany}
\author{T.~G\"orler} \affiliation{Max Planck Institute for Plasma Physics Boltzmannstr 2 85748 Garching Germany}
\author{E.~Poli} \affiliation{Max Planck Institute for Plasma Physics Boltzmannstr 2 85748 Garching Germany}
\author{A.~Ba\~n\'on~Navarro} \affiliation{Max Planck Institute for Plasma Physics Boltzmannstr 2 85748 Garching Germany}
\author{A.~Biancalani} \affiliation{Max Planck Institute for Plasma Physics Boltzmannstr 2 85748 Garching Germany}
\author{F.~Jenko} \affiliation{Max Planck Institute for Plasma Physics Boltzmannstr 2 85748 Garching Germany}

\date{\today}

\begin{abstract}

In recent years, a strong reduction of plasma turbulence in the presence of energetic particles has been reported in a number of magnetic confinement experiments and corresponding gyrokinetic simulations. While highly relevant to performance predictions for burning plasmas, an explanation for this primarily nonlinear effect has remained elusive so far. A thorough analysis finds that linearly marginally stable energetic particle driven modes are excited nonlinearly, depleting the energy content of the turbulence and acting as an additional catalyst for energy transfer to zonal modes (the dominant turbulence saturation channel). Respective signatures are found in a number of simulations for different JET and ASDEX Upgrade discharges with reduced transport levels attributed to energetic ion effects. 

\end{abstract}


\maketitle

{\em Introduction.} Being an almost ubiquitous phenomenon, turbulence with its highly stochastic and nonlinear character is a subject of active research in various fields. In magnetically confined plasma physics, it is of particular interest since it largely determines the radial heat and particle transport and thus the overall confinement. Any insight on possible reductions of the underlying micro-instabilities which are driven by the steep density and temperature profiles, and/or on modifications of their nonlinear saturation mechanisms can be considered crucial on the way to self-sustained plasma burning and corresponding fusion power plants. A particularly interesting example is the recent experimental and numerical evidence suggesting a link between the presence of fast ions and substantial improvement of energy confinement in predominantly ITG (ion-temperature-gradient) driven turbulence \cite{Tardini_NF2007,MRomanelli_PPCF2010,Holland_NF2012,Citrin_PPCF2015,Citrin_PRL2013}. Dedicated theoretical studies have already identified a number of possible energetic ion effects on plasma turbulence like dilution of the main ion species \cite{Tardini_NF2007}, Shafranov shift stabilization \cite{Bourdelle} and resonance interaction with bulk species micro-instabilities in certain plasma regimes \cite{DiSiena_NF2018}. They furthermore contribute to the total plasma pressure and increase the kinetic-to-magnetic pressure ratio, $\beta$, which is a measure for the relevance of electromagnetic fluctuations, known to stabilize ITG modes. Such behaviour could indeed be confirmed in simulations \cite{Citrin_PPCF2015,Garcia_NF2015,Doerk_NF2018} of JET hybrid discharges \cite{Mantica_PRL2009,Mantica_PRL2011} with substantial fast ion effects that, however, also identified an upper limit for this beneficial fast-ion-pressure effect. If the total plasma pressure exceeds a critical value, kinetic ballooning or Alfv\'enic ITG modes with smaller toroidal mode numbers and frequencies higher than the ITG modes are destabilized which increase particle/heat fluxes \cite{Pueschel_2010}. Similarly, the fast-ion pressure may drive energetic particle (EP) modes if certain thresholds are exceeded. Although a possible relevance of the proximity to the onset of these modes has been noted \cite{Citrin_PPCF2015,Garcia_NF2015}, their role was not investigated in more detail.
In any case, all of these effects are mainly linear, i.e., alter the growth of the underlying micro-instability. A satisfactory explanation for the particularly strong nonlinear reduction in electromagnetic simulations with fast ions \cite{Citrin_PRL2013,Doerk_NF2018} still represents an outstanding issue. 
A substantially stronger nonlinear transport reduction compared to linear simulations is also found in electromagnetic studies without fast ions. Here, a theoretical model \cite{Whelan_PRL2018} has recently been proposed which suggests that electromagnetic fluctuations strengthen the nonlinear interaction between a large variety of unspecified stable and unstable modes with zonal flows (ZF) by increasing the so-called triplet correlation time. The further enhancement of this effect by fast ions was, however, not covered.
   
This letter is therefore dedicated to fill the missing gaps and to provide - for the first time - a consistent picture of the nonlinear impact of fast ions on plasma turbulence. For this purpose, gyrokinetic simulation results for one specific scenario, namely the JET-like scenario described below, will be analyzed in detail with a new approach of frequency-spectral decomposition of the free energy balance. This allows the identification of the effect by which nonlinearly excited modes catalyze the main saturation mechanism and therefore substantially decrease the turbulent transport levels. All the results presented in this letter are also observed in a number of simulations for ASDEX Upgrade and JET discharges where the improved ion energy confinement was attributed to nonlinear electromagnetic energetic ion effects ~\cite{Citrin_PPCF2015,Citrin_PRL2013,Doerk_NF2018,Bonanomi_NF_2018}.

\vspace{3mm}

\begin{figure}
\begin{center}
\includegraphics[scale=0.32]{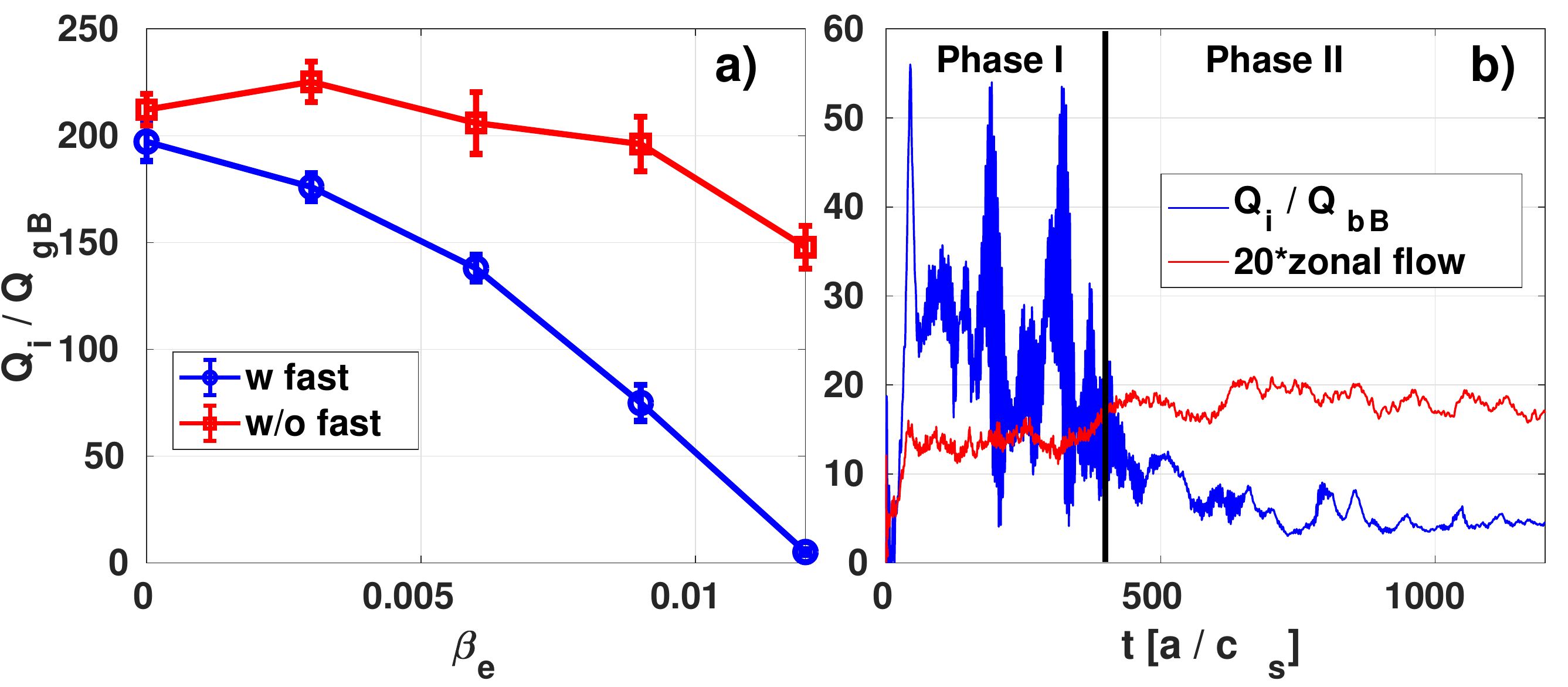}
\par\end{center}
\caption{Nonlinear main ion heat flux in GyroBohm units for a) different $\beta_e$ and b) time trace and zonal amplitude at $\beta_e = 0.012$ with fast ions. The vertical black line marks the time of the transition between phase I and II.}
\label{fig:fig1}
\end{figure}

\begin{figure}
\begin{center}
\includegraphics[scale=0.32]{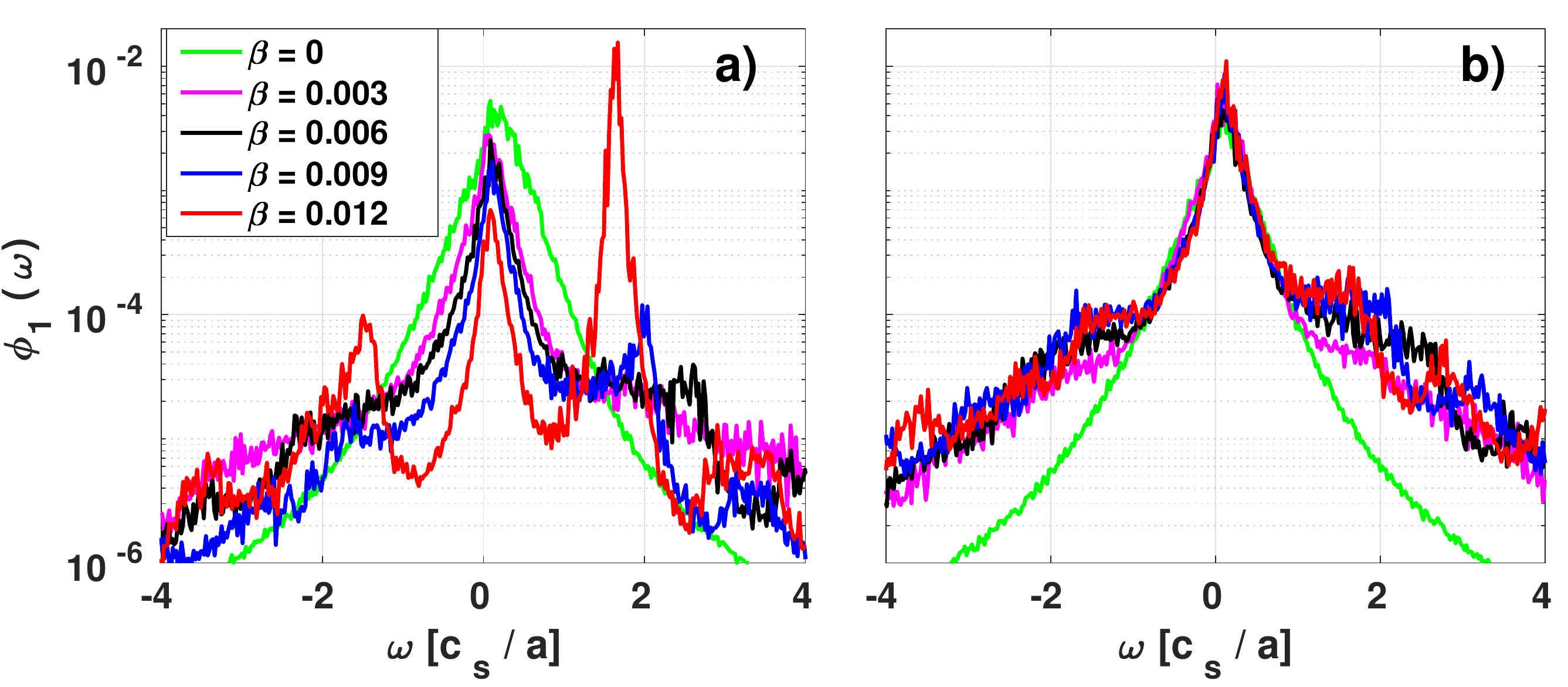}
\par\end{center}
\caption{Frequency spectra of $\bar{\phi}_1$ - averaged over $k_x \rho_s$ and $z$ - for different $\beta_e$ at $k_y \rho_s = 0.1$ for simulations a) with b) without fast ions in the time range $[50 - 340] a / c_s$. The plots share the same legend. The ITG-frequency peaks at $\omega/[c_s / a] \sim 0.08$, while the TAE frequency follows the relation $\omega / [c_s / a] = v_{th,i} / (2 q R_0 \sqrt{\beta_i})$ for each value of $\beta_e$.}
\label{fig:fig2}
\end{figure}

{\em Simulations: Setup and Results.}
The impact of the fast ions on the electromagnetic nonlinear stabilization of plasma turbulence is investigated with GENE \cite{Jenko_PoP2000} turbulence flux-tube simulations. Our study is based on the JET-like scenario with deuterium, electrons and externally injected neutral beam deuterium described in detail in Ref.~\cite{DiSiena_NF2018,Bravenec_PPCF2016}, with a reduced safety factor of $q = 1.2$. The grid resolution in radial, binormal and parallel to the magnetic field line directions is $(x,y,z) = (192,96,32)$ points, while in the magnetic moment and parallel velocity $(\mu,v_\shortparallel) = (20,32)$ points. The radial box size is $175 \rho_s$ and the minimum $k_y \rho_s = 0.025$ with thermal gyroradius $\rho_s = (T_e / m_i)^{1/2} / \Omega_i$. Here, $\Omega_i = m_i c / q B_0$ denotes the gyro-frequency, $T_i$ the main ion temperature, $m_i$ the main ion mass, $q$ the charge, $B_0$ the magnetic field on axis and $c$ is the speed of light. 
The basic finding to be explained is displayed in Fig.~\ref{fig:fig1}a) where the flux surface averaged heat fluxes -- normalized to $Q_{gB} = v_{th,i}\rho_i^2n_eT_i/R_0^2$ -- are shown for different values of the electron thermal to magnetic pressure ratio $\beta_e = 8\pi p_e /B_0^2$ in simulations with/without energetic particles.
By including the energetic ion species, a pronounced reduction in the turbulent fluxes with $\beta_e$ is observed compared to the electrostatic case - ca.~$97\%$ at $\beta_e = 0.012$, which by far exceeds the analogous linear growth rate stabilization - ca.~$30\%$. Moreover, the presence of fast ions yields a substantial further stabilization compared to electromagnetic simulations without, e.g.~$95\%$ at $\beta_e = 0.012$. Considering the heat flux time traces of the higher $\beta_e$ simulations with strong transport reduction in more detail, two nonlinear phases can usually be observed as shown in Fig.~\ref{fig:fig1}b). 
A striking observation during the first phase are high-frequency modulations of the heat fluxes in the presence of fast ions. They can be attributed to {\em linearly marginally stable} EP-driven modes. Further analysis shows that they are found to lie at the center of the SAW (shear Alfv\'en wave) toroidicity-induced gap \cite{Cheng_PF_1985,Chen_RMP_2016}, exhibit the TAE (toroidal-Alfv\'en-eigenmodes) frequency $\omega / [c_s / a] = v_{th,i} / (2 q R_0 \sqrt{\beta_i})$ for each value of $\beta_e$ and become unstable at $\beta_e\sim 0.013$. $R_0$ is expressed in units of the minor radius.
Further signatures for their presence are given in Fig.~\ref{fig:fig2}a) where a Fourier transform has been applied to the  gyroaveraged electrostatic potential $\bar{\phi}_1$ in the first nearly-steady state time range (phase I). Clearly, a progressive destabilization of high-frequency components with $\beta_e$ can be seen at $k_y \rho_s = 0.1$ while no such significant difference in the electrostatic potential frequency spectra can be observed in the absence of EP in Fig.~\ref{fig:fig2}b). Moreover, Fig~\ref{fig:fig2}a) shows a corresponding reduction of the ITG-frequency domain peak $(\omega/[c_s / a] \sim 0.08)$ from ca.~$30\%$ for $\beta_e = 0.003$ to ca.~$85 \%$ for $\beta_e = 0.012$ with respect to the electrostatic limit, as the high-frequency mode is destabilized. 
Considering all $k_y$ wave numbers, the presence of this mode is observed in a wider spectral ($k_y$) range which broadens with increasing $\beta_e$. For the case $\beta_e = 0.012$, for instance, high-frequency fluctuations are observed up to ITG relevant binormal mode-numbers, namely $0.025 < k_y \rho_s < 0.2$ with a maximum at $k_y \rho_s = 0.15$. 
During the first nonlinear phase, the energy enclosed in the TAE frequency range, namely $1.3 < \omega / [c_s / a] < 2.5$, increases from ca.~$ 0\%$ to ca.~$30\%$ as $\beta_e$ is varied from 0 to $0.012$, with a reduction in the ITG-frequency free energy content. These results are consistent with the amplitude reduction of the electrostatic potential of the ITG peak due to nonlinear coupling to the TAE mode discussed below. In this phase, the zonal flow levels seem to be hardly affected by the EP presence such that the overall EP induced transport reduction remains moderate. On the contrary, during the second nonlinear phase, a significant increase in zonal component of the potential is observed, substantially reducing ion-scale turbulence transport.

\vspace{3mm}
{\em Nonlinear energy-transfer analysis.} In order to understand the phenomenology described above, the energy transfers shall be studied more closely, e.g., by monitoring the nonlinear mode-to-mode coupling term in the free energy balance equation \cite{Nakata_PoP2012,Maeyama_PRL2015,Maeyama_PRL2017}
\begin{eqnarray*}
\mathcal{N}_{k} = \sum_{k^\prime,k^{\prime\prime}} \mathcal{T}_{k}^{k^\prime,k^{\prime\prime}}=\sum_{s,k^\prime,k^{\prime\prime}} \Re e\Bigg\{ \int dz dv_\shortparallel d\mu \pi B_0 h_{1,s}^{k,*}
\end{eqnarray*}
\begin{eqnarray}
\frac{n_s T_s}{F_{0,s}} \left[ \left( \vec{k}^{\prime} \times \vec{k}^{\prime\prime} \right) \cdot \frac{\vec{B}_0}{|B_0|} \right] \left( \bar{\xi}_{1,s}^{k^\prime} g_{1,s}^{k^{\prime\prime}} - \bar{\xi}_{1,s}^{k^{\prime\prime}} g_{1,s}^{k^{\prime}} \right) \Bigg\},
\label{eq:eq2}
\end{eqnarray}
with $h_{1,s}^k =f_{1,s}^k + q_s \bar{\phi}_{1}^k F_{0,s}/T_s$. Here, $s$ denotes the plasma species with density $n_s$, temperature $T_s$, and charge $q_s$. Furthermore, $F_{0,s}$ represents the Maxwellian background, and $g_{1,s} = f_{1,s} + q_s v_{th,s} v_\shortparallel F_{0,s} \bar{A}_{1,\shortparallel} / T_s$ a modified distribution with the perturbed distribution function $f_{1,s}$, the thermal velocity $v_{th,s} = \sqrt{2 T_s / m_s}$, the gyroaveraged parallel component of the vector potential $\bar{A}_{1,\shortparallel}$ and the field $\bar{\xi}_{1,s} = \bar{\phi}_1 - v_{th,s} v_\shortparallel \bar{A}_{1,\shortparallel}$. The symbol $\mathcal{T}_{k}^{k^\prime,k^{\prime\prime}}$ represents the nonlinear energy transfer between the modes $k$, $k^\prime$ and $k^{\prime\prime}$. It is a cubic function of $g_{1,k}$ and it can be expressed as a triadic nonlinear coupling between the modes $k$, $k^\prime$ and $k^{\prime\prime}$. Since the coupling condition $k + k^\prime + k^{\prime\prime} = 0$ is satisfied, the triad transfer is a symmetric function of $k^\prime$ and $k^{\prime\prime}$, i.e.~$\mathcal{T}_{k}^{k^\prime,k^{\prime\prime}} = \mathcal{T}_{k}^{k^{\prime\prime},k^\prime,}$. In the gyrokinetic formalism, the free energy is a nonlinearly conserved quantity, i.e.~$\mathcal{N}_{k}$ vanishes when summed over all the wave vector components. Thus, $\mathcal{T}_{k}^{k^\prime,k^{\prime\prime}}$ represents the transfer of energy from/to different scales.
\begin{figure}
\begin{center}
\includegraphics[scale=0.32]{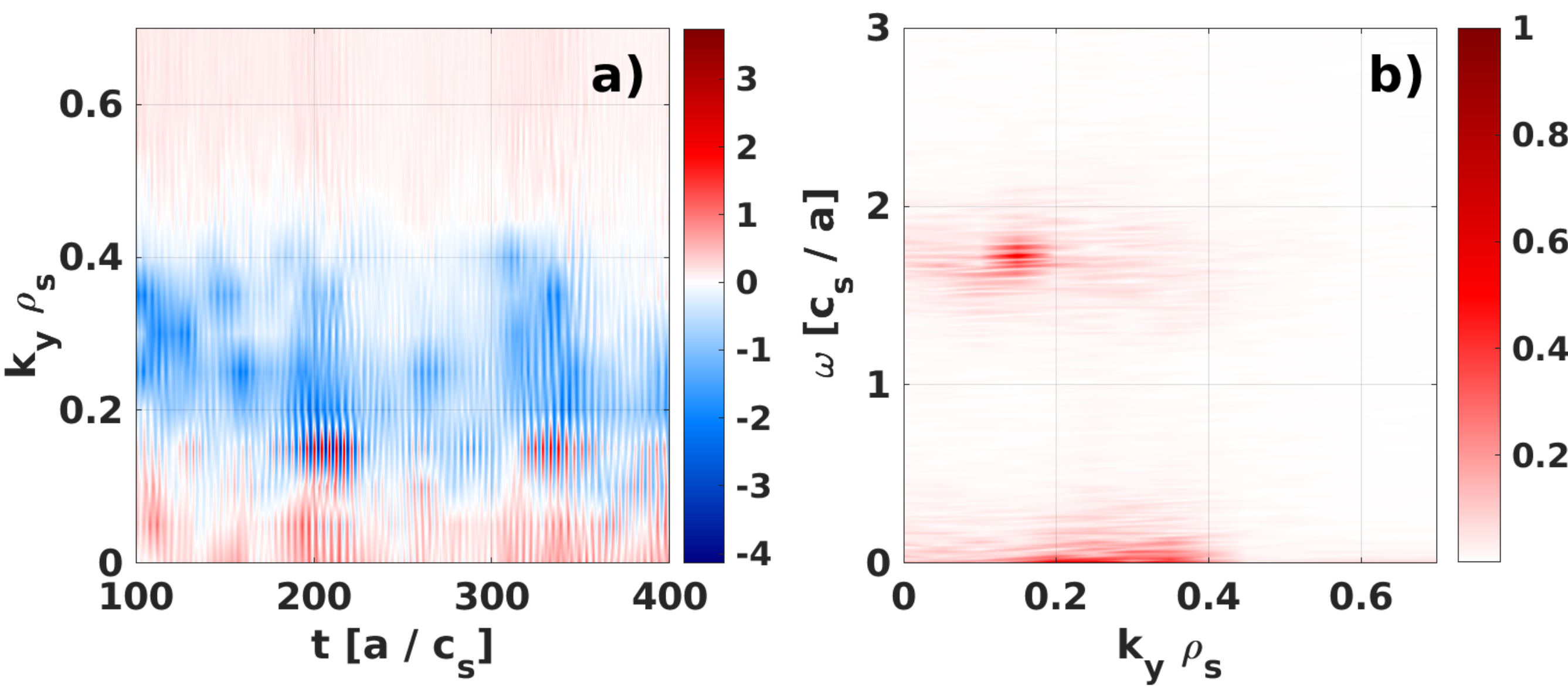}
\par\end{center}
\caption{Nonlinear transfer at $\beta_e = 0.012$ - averaged over $k_x \rho_s$ and $z$ - in $k_y \rho_s$ and a) in the time domain $[100 - 400] a/ c_s$ and its b) frequency spectra. Positive and negative values mean, respectively, that the given wave-vector is receiving or losing energy through nonlinear coupling.}
\label{fig:fig3}
\end{figure}
Fig.~\ref{fig:fig3}a) shows the time evolution of the nonlinear mode-to-mode coupling term $\mathcal{N}_{k_y}\left(t\right)$ in the first nonlinear phase, summed over all radial wave-numbers for the simulation at $\beta_e = 0.012$. A significant energy transfer is observed from the ITG-relevant binormal wave-vectors $0.2 < k_y \rho_s < 0.45$ to larger scales $0 < k_y \rho_s < 0.175$. Although its structure is not affected by the amplitude of the magnetic fluctuations, the nonlinear energy exchange rate $\mathcal{N}\left(\omega\right)$ significantly increases with $\beta_e$. By performing a Fourier decomposition in time of $\mathcal{N}$ for each $k_y \rho_s$, fast oscillations are observed in Fig.~\ref{fig:fig3}b) for the binormal wave vector range $0.025 < k_y \rho_s < 0.2$ at the specific TAE mode frequency. The mode-to-mode coupling term transfers energy from ITG- to TAE-scales and is strongly enhanced by $\beta_e$. As observed before, the energy content in the
spectral region corresponding to the TAE increases to ca 30\% for $\beta_e = 0.012$. These results are consistent with the frequency peaking of the electrostatic potential of Fig.~\ref{fig:fig2}a). 

\vspace{3mm}

\begin{figure}
\begin{center}
\includegraphics[scale=0.32]{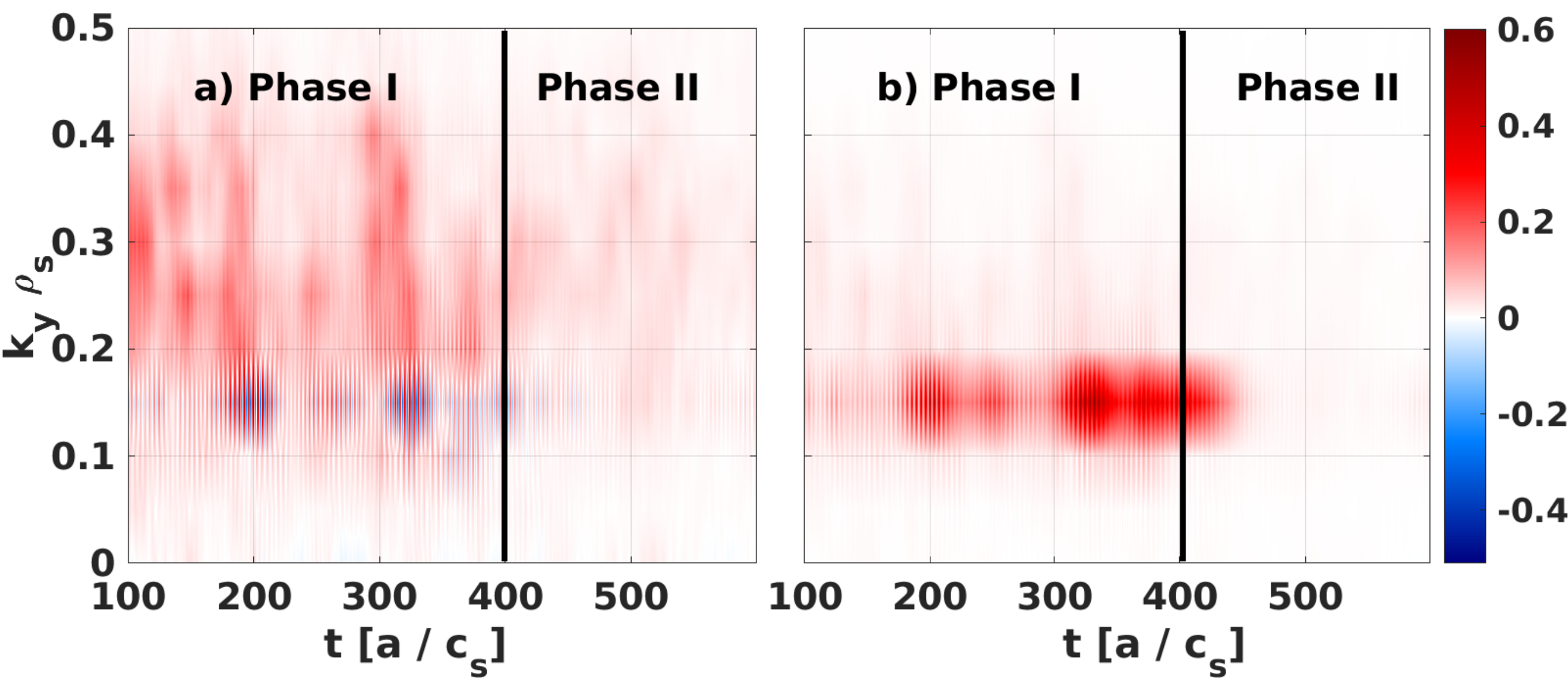}
\par\end{center}
\caption{Time trace of the field component of the curvature term in the free energy balance for a) thermal deuterium and b) NBI - averaged over $k_x \rho_s$ and z - for different $k_y \rho_s$ at $\beta_e = 0.012$. Positive/negative values indicate a destabilizing/stabilising contribution to the overall drive at the given wave number. The black lines mark the time of the transition between phase I and II.}
\label{fig:fig4}
\end{figure}

It was noted previously that linear energy balance analyses at $k_y \rho_s = 0.1$ show that the EP-driven mode is linearly stable for $\beta_e < 0.013$, being suppressed by Landau damping mechanisms. However, as $\beta_e$ increases, the curvature term contribution to the linear instability increases significantly, with a reduction of the linear damping from $\gamma_{TAE} = -0.124 c_s /a$ at $\beta_e = 0.003$ to $\gamma_{TAE} = -0.005 c_s / a $ at $\beta_e = 0.012$. As this mode becomes closer to the marginal stability, more and more energy is exchanged nonlinearly with the dominant ion-scale turbulence through mode-to-mode coupling. The interplay between nonlinear drive and damping of the EP-driven mode can be studied in detail by investigating the field component of the free energy balance \cite{Banon_Navarro_PoP2011}
\begin{equation}
    \frac{\partial E_{w}^k}{\partial t} = \sum_s \Re e\left\{ \int dz dv_\shortparallel d\mu \pi B_0 n_s q_s \bar{\phi}_{1}^{k,*} \frac{\partial g_{1,s}^k}{\partial t} \right\}.
    \label{eq:eq1}
\end{equation}
This analysis is reduced to the study of the curvature term - usually destabilizing - and parallel advection - related to Landau damping mechanisms. Fig.~\ref{fig:fig4} reveals that, during the first phase, significant energy is transferred from the main deuterium to the EP curvature term, which reaches amplitudes similar to the thermal species. This interaction, identified by the oscillatory pattern of Fig.~\ref{fig:fig4}, occurs at the TAE scale, namely $k_y \rho_s \sim 0.15$ and is modulated at the TAE frequency. Moreover, Fig.~\ref{fig:fig4} shows that EPs provide the dominant contribution to the high-frequency mode, consistently with the lack of turbulence stabilization observed in their absence in Fig~\ref{fig:fig1}a). These results explain the progressive stabilization observed in the first phases of the nonlinear simulations with $\beta_e$. 
In correspondence with the second nonlinear phase, the amplitude of the main deuterium curvature term decreases significantly with non-negligible EP contributions. The latter, however, sustained only through nonlinear coupling with ITG-scales, drops at a later time - $t \sim 430 a/c_s$ - as well, as a consequence of the lack of cross-scale transferred energy.

\vspace{3mm}

\begin{figure}
\begin{center}
\includegraphics[scale=0.52]{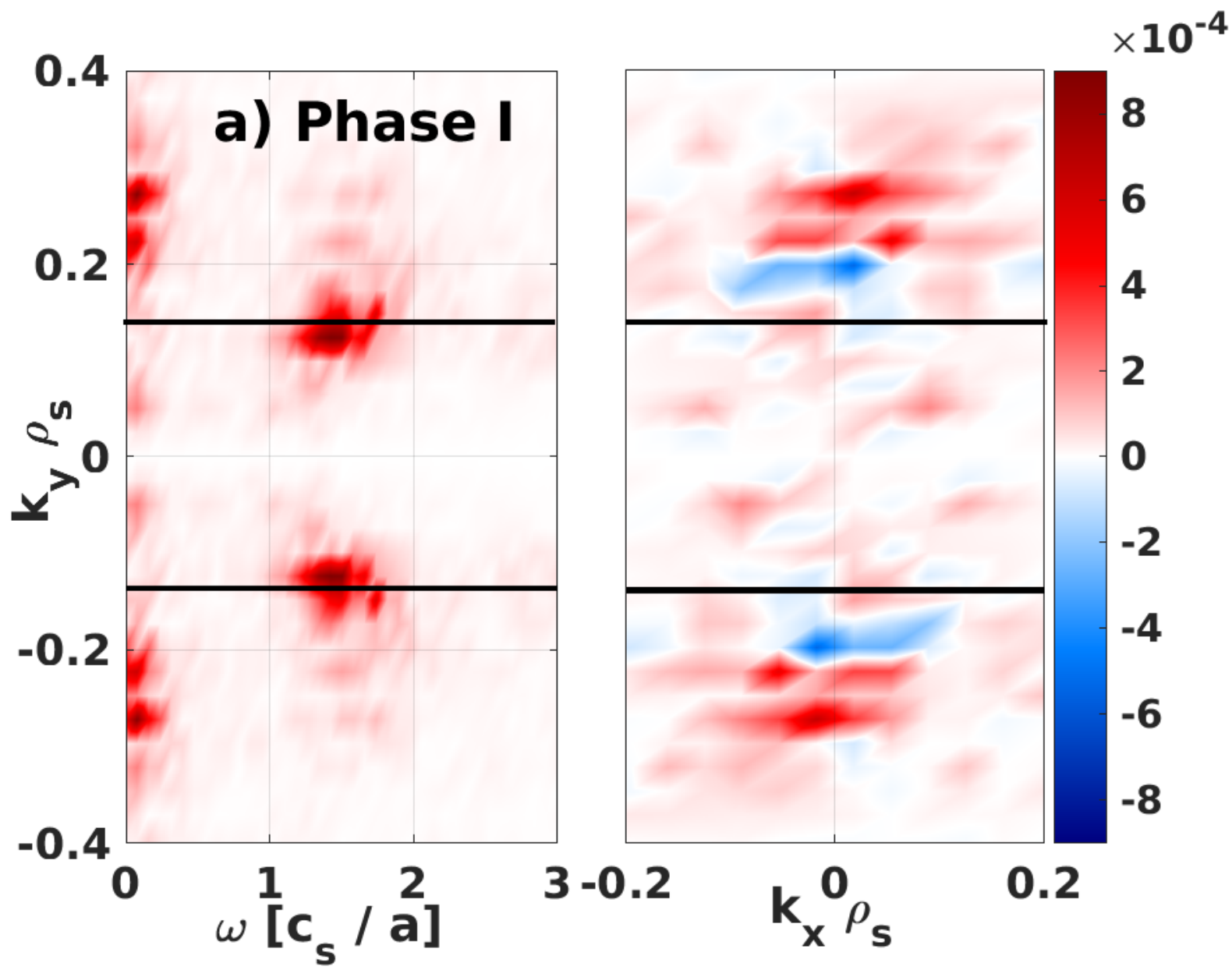}
\includegraphics[scale=0.52]{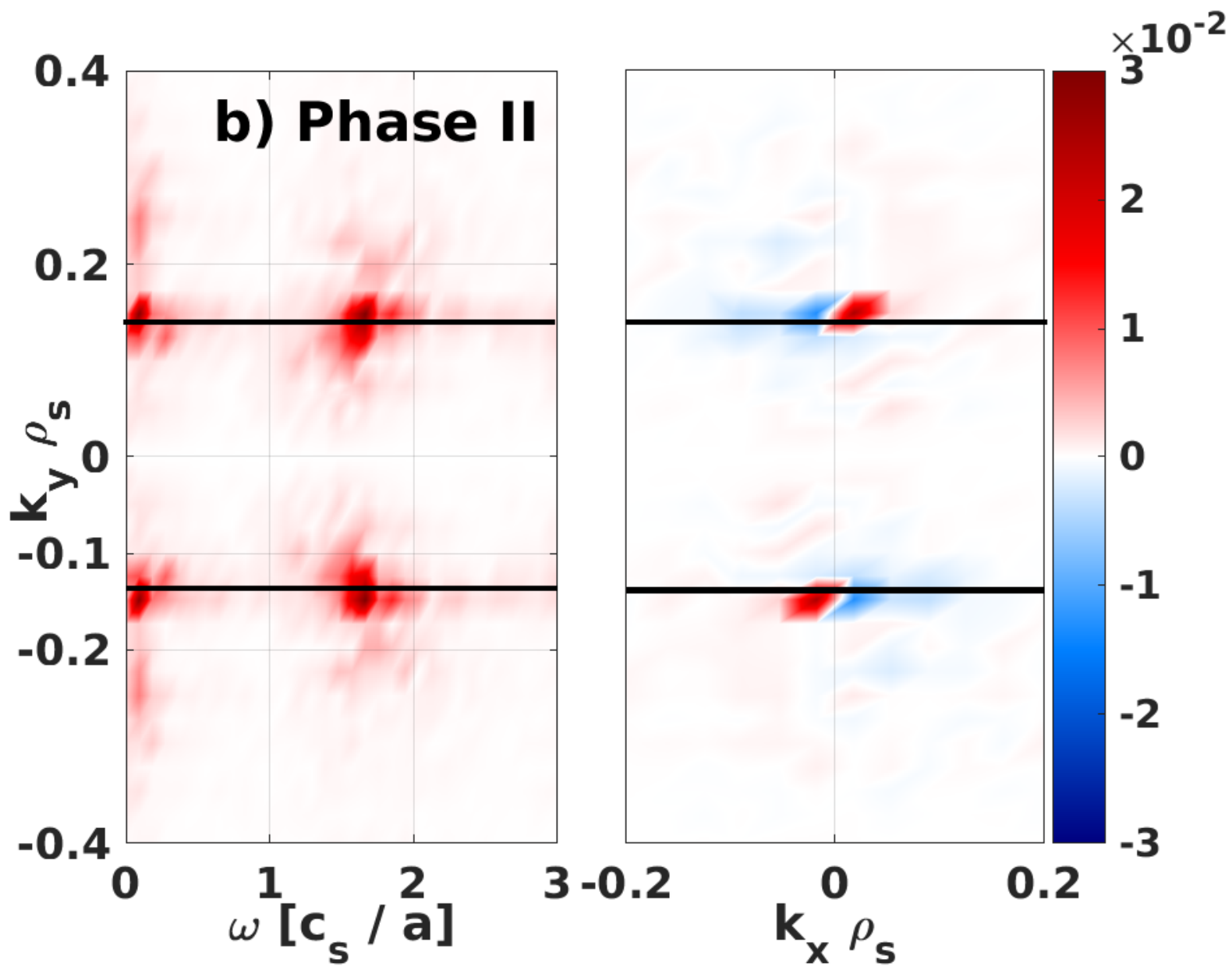}
\par\end{center}
\caption{Triad energy transfer normalized to the main deuterium heat flux as function of $(\omega,k_y)$ deep in the a) first $(t=165-245\,a/c_s)$ and b) second phase $(t=470-550\,a/c_s)$. The black line denotes the dominant TAE scale $k_y \rho_s = 0.15$. Positive/negative values indicate that a given wave-vector is receiving/losing energy through nonlinear coupling with the resonant modes. Note the difference in the colorbar, i.e.~in the amplitudes in the two phases.}
\label{fig:fig5}
\end{figure}

The transition between the two nonlinear phases occurs with an increase of the ZF levels as shown in Fig.~\ref{fig:fig1}b) for $\beta_e = 0.012$. The TAE starts to interact consistently with the zonal field component after reaching amplitudes similar to those of the thermal drive (as shown in Fig.~\ref{fig:fig4}). The "triad" coupling function $\mathcal{T}_{k}^{k^\prime,k^{\prime\prime}}$, defined in Eq.~\ref{eq:eq2}, is employed to investigate in detail the difference in the nonlinear interaction between EP-driven TAE and ZF in the two phases of the nonlinear simulations with energetic particles. Fig.~\ref{fig:fig5} shows the triad wave-number spectra, normalized to the main ion heat flux, averaged over the time domains of the two phases for $\left(k_x, k_y \right)\rho_s = \left(0.04, 0 \right)$, i.e.~for transfer to the zonal component. No significant difference is found if the radial wave-number $k_x$ is changed. In the first phase, the selected triplet is interacting mainly with the binormal mode numbers in the range $0.2 < k_y \rho_s < 0.4$, as can be seen in Fig.~\ref{fig:fig5}a). At this scale, the ITG-drive peaks and the time-averaged EP-driven mode contribution is negligible, as confirmed by the frequency decomposition of the time trace of the overall triplet. In the first nonlinear phase, the TAE mode is not interacting significantly with zonal modes. However, as the energy is nonlinearly transferred from ITG to TAE scales, the amplitude of the EP-driven modes increases significantly and ZFs are more and more affected by the presence of these modes. In the second nonlinear phase, the whole energy transfer to the specific triplet occurs through the wave-vector $k_y \rho_s = 0.15$, where the TAE mode is dominant and it overcomes the thermal ITG contribution. The energy exchange increases by a factor of $\sim 30$. 
The TAE mode therefore acts as an additional mediator of plasma turbulence, catalyzing energy transfer to zonal modes, strongly affecting the standard paradigm of ZF/ion-scale-turbulence interaction \cite{Nakata_PoP2012, Diamond_PPCF2005}. The physical mechanism described in this letter may very well open ways for new physical interpretations of more general turbulent systems well beyond the scenarios which involve energetic ions and magnetically confined plasmas. In particular, similar nonlinear effects might be observed each time subdominant modes approach the marginal stability threshold and are allowed to couple with both the dominant instabilities and the stable modes acting as main saturation players. Particularly strong nonlinear reductions have, e.g., also been found in the absence of energetic ions in transitions from trapped-electron to ITG modes \cite{Merz_NF_2010} and are an obvious subject for further investigations along these lines. Furthermore, the above described mediator effect may also be interesting to neighbouring communities such as optics \cite{Lin_IJE_1992, Wang_PRL_1995}, fluid dynamics \cite{Saarloos_PRA_1989} and field theory \cite{Cai_PRD_2011, Gubser_PRD_2008}.

\vspace{3mm}
{\em Conclusions.}
The intriguing and particular strong transport reduction in the presence of fast ions observed in several scenarios could - for the first time - be explained by their ability to trigger marginally stable modes which are nonlinearly excited and act as a catalyst for the main turbulence saturation mechanisms. While already being highly relevant to plasma physics with strong heating, this study may furthermore motivate deliberate "design" of marginally stable modes in order to exploit their capability as mediators boosting nonlinear saturation mechanisms such as zonal flows. 

In the case discussed in this paper -- a strongly NBI-heated JET discharge with fast ion related temperature profile steepening -- a two-phase process could be observed in nonlinear gyrokinetic simulations and analyzed with new spectral analyses techniques. The fast ions provide linearly marginally stable TAE modes which are nonlinearly excited by an energy redistribution from ITG to TAE spatio-temporal scales. As a result, lower transport levels corresponding to the net reduction of the ITG drive can be observed. If sufficiently populated, the fast ion modes furthermore start to increasingly affect the ZF levels which marks a second phase in the simulations. During this phase, the energy being nonlinearly transferred to zonal modes undergoes a substantial increase in magnitude and is modulated at the TAE frequency. The increase in ZF levels directly impacts the ion-scale turbulence, strongly suppressing heat/particle fluxes. This in turns lowers the nonlinear drive of the EP modes. The system finally finds an equilibrium at a much reduced transport level. This mechanism with possibly high relevance to future plasma performance predictions is not restricted to the scenario analyzed here, but could also be identified in different JET and ASDEX Upgrade discharges with strong heating. 

The simulations presented in this work were performed at the Cobra HPC system at the Max Planck Computing and Data Facility (MPCDF), Germany. Furthermore, we acknowledge the CINECA award under the ISCRA initiative, for the availability of high performance computing resources and support. The authors would like to thank N.~Bonanomi, J.~Citrin, H.~Doerk, P.~Lauber, P.~Manas, P.~Mantica, I.~Novikau, K.~Stimmel and A.~Zocco for all the stimulating discussions, useful suggestions and comments.

\end{document}